\documentclass[12pt]{amsart}
\usepackage{amssymb,amsfonts,latexsym,amscd}

\def\1{{\bf 1}}
\def\a{\alpha}
\def\al{a_\lambda}

\def\d{{\rm d}}

\def\Re{{\mathcal Re}}

\def\ua{\uparrow} 
\def\da{\downarrow} 
\def\cw{C_W}

\def\bra{\langle}

\def\ket{\rangle}

\def\C{{\mathbb{C}}} 
\def\R{{\mathbb{R}}}  

\def\gm{{\gamma}}

\newcommand{\gH}{{\mathfrak H}}  
\newcommand{\gF}{{\mathfrak F}}  

\def\eqnn{\begin{eqnarray*}}
\def\eeqnn{\end{eqnarray*}}
\def\eqn{\begin{eqnarray}}
\def\eeqn{\end{eqnarray}}
\def\bal{\begin{align}}
\def\eal{\end{align}}

\newtheorem{theorem}{Theorem}[section]

\newtheorem{corollary}[theorem]{Corollary}
\newtheorem{lemma}[theorem]{Lemma}

\newtheorem{proposition}[theorem]{Proposition}
\newtheorem{remark}[theorem]{Remark}

\begin{document}

\title[Quantitative estimates on the enhanced binding]
{Quantitative estimates on the enhanced binding
for the Pauli-Fierz operator}

\thanks{The authors gratefully acknowledge financial
support from the following institutions: The European Union
through the IHP network ``Analysis and Quantum''
HPRN-CT-2002-00277 (JMB and SV), the French Ministery of Research
through the ACI ``jeunes chercheurs" (JMB), the Volkswagen
Stiftung and DIPUC of the Pontific\'\i a Universidad Cat\'olica
(HL), and the DFG grant WE 1964/2 (SV)}

\author{Jean-Marie Barbaroux \and Helmut Linde
\and Semjon Vugalter}

\date{10 august 2005}
\address{Centre de Physique Th\'eorique, Luminy Case 907, 13288
Marseille Cedex~9, France and D\'epartement de Math\'ematiques,
Universit\'e du Sud-Toulon-Var, avenue de l'Universit\'e, 83957 La
Garde Cedex, France, barbarou@univ-tln.fr}
\address{Facultad de F\`isica,
P. Universidad Cat\'olica de Chile, Casilla 306, Santiago 22,
Chile }
\address{Mathematisches Institut, Ludwig-Maximilians-Universit\"at
M\"unchen, Theresienstrasse 39, 80333 M\"unchen and Institut f\"ur
Analysis, Dynamik und Modelierung, Universit\"at Stuttgart,
wugalter@mathematik.uni-muenchen.de}

\maketitle
\begin{abstract}
For a quantum particle interacting with a short-range potential,
we estimate from below the shift of its binding threshold, which
is due to the particle interaction with a quantized radiation
field.
\end{abstract}


\section{Introduction}


Recently, the question of enhanced binding in nonrelativistic QED
has been extensively studied in several publications
\cite{HiroshimaSpohn2001, Hainzletal2003, Chenetal2003,
CattoHainzl2004, BenguriaVugalter2004, Cattoetal2004}. Dressing a
charged particle with photons increases the ability of a potential
to confine it.
For the Pauli-Fierz operator which describes a nonrelativistic particle interacting with a radiation field, this effect
was proved for small values of the fine structure constant $\alpha$, first under the simplifying assumption that the
spin of the particle is absent \cite{Hainzletal2003}, and later generalized to the case of a particle with spin
\cite{Chenetal2003, CattoHainzl2004}. In \cite{BenguriaVugalter2004}, it was shown that the effect of the enhanced
binding is asymptotically small in $\alpha$ in the sense that the binding threshold for the Pauli-Fierz operator tends
to the binding threshold for the corresponding Schr\"odinger operator as $\alpha$ tends to zero. Some quantitative
estimates on this effect were obtained in \cite{Cattoetal2004} where it was proved that the difference between the
binding threshold for the Schr\"odinger operator and the corresponding Pauli-Fierz operator with spin zero is at least
of the order $\alpha$. In the work at hand, using a different method, we prove similar results for the more general
case of a particle with spin zero or one half. Notice that studying the enhanced binding effect in the case of a
particle with spin requires recovering one more term of the energy's expansion in powers of $\alpha$ than in the
spinless case.

The method of the proof is a further development of a method used in \cite{Hainzletal2003} and \cite{Chenetal2003}. We
prove that the Pauli-Fierz operator has a ground state even for some value of the potential coupling constant that is
smaller than the binding threshold for the corresponding Schr\"odinger operator. To do so, we construct a trial
function for which the quadratic form of the Pauli-Fierz operator with this coupling constant takes a value strictly
less than the self-energy. Then we apply \cite[Theorem~2.1]{Griesemeretal2001} which tells us that this implies the
existence of a ground state. The trial function we use is similar to the one in \cite{Chenetal2003} with some
modifications necessary to obtain quantitative estimates in the case with spin. It is constructed using the ground
state of the self-energy operator with total momentum zero.

As in all previous papers \cite{Hainzletal2003, Chenetal2003,
CattoHainzl2004, BenguriaVugalter2004, Cattoetal2004}, our method
is asymptotic in $\alpha$. Therefore, the problem of establishing
the enhanced binding effect and estimating its strength for the
physical value of $\alpha\approx 1/137$ still remains open.

\section{Definitions and main result}\label{S1}

The Pauli-Fierz Hamiltonian $H$ for a charged particle with or without spin in an external electrostatic potential and
coupled to the quantized electromagnetic radiation field is defined by
\begin{equation}\label{rpf}
\begin{split}
H \!\!=\!\!  \left(- i\nabla_{x}\otimes I_f\! +\! \sqrt{\alpha}
A(x)\right)^2\!\! +\! g\sqrt{\alpha}\sigma\!\!\cdot\!\! B(x)\!\!
+\!\! \lambda W(x)\!\otimes\! I_f\! +\! I_{el}\!\otimes\! H_f\!
-\!c_{\rm n.o.}\alpha.
\end{split}
\end{equation}
The operator $H$ acts on the Hilbert space $\gH:=\gH^{el}\otimes \gF$. The Hilbert space $\gH^{el}$ of the
nonrelativistic particle is $L^2(\R^3)\otimes \C^2$ in the case $g=1$ and $L^2(\R^3)$ in the case $g=0$. Here $\R^3$ is
the configuration space of a single particle, while $\C^2$ accommodates its spin in the case $g=1$.

We will describe the quantized electromagnetic field by use of the
Coulomb gauge condition. Accordingly, the one-photon Hilbert space
is given by $L^2(\R^3)\otimes \C^2$, where $\R^3$ denotes either
the photon momentum or configuration space, and $\C^2$ accounts
for the two independent transversal polarizations of the photon.
The photon Fock space is then defined by
 $$
   \gF = \bigoplus_{n=0}^\infty \gF_s^{(n)} ,
 $$
where the n-photons space $\gF_s^{(n)} =
\bigotimes_s^n\left(L^2(\R^3)\otimes\C^2\right)$ is the symmetric
tensor product of $n$ copies of $L^2(\R^3)\otimes\C^2$.

We use units such that $\hbar = c = 1$, and where the mass of the
particle equals $m=1/2$. The particle charge is then given by
$e=\sqrt{\alpha}$. As usual, we will consider $\alpha$ as a small
parameter.

The operator that couples a particle to the quantized vector
potential is given by
\begin{equation}\nonumber
\begin{split}  A(x) = & \sum_{\lambda = 1,2} \int_{\R^3}
  \frac{\zeta(|k|)}{2\pi|k|^{1/2}}
  \varepsilon_\lambda(k)\Big[ e^{ikx} \otimes a_\lambda(k)  +
  e^{-ikx} \otimes a_\lambda^\ast
  (k) \Big] \d k \\
  =: &D(x) + D^\ast (x),
\end{split}
\end{equation}
where ${\rm div}A =0$ by the Coulomb gauge condition. The operators $a_\lambda$, $a_\lambda^*$ satisfy the usual
commutation relations
 $$
  [a_\nu(k), a^\ast_\lambda(k')] = \delta (k-k') \delta_{\lambda, \nu},
  \quad [a_\nu(k), a_\lambda(k')] = 0 .
 $$
The vectors $\varepsilon_\lambda(k)\in\R^3$ are the two
orthonormal polarization vectors perpendicular to $k$,
\begin{equation}\label{def-eps}
  \varepsilon_1(k) = \frac{(k_2, -k_1, 0)}{\sqrt{k_1^2 + k_2^2}}\qquad
 {\rm and} \qquad
   \varepsilon_2(k) = \frac{k}{|k|}\wedge \varepsilon_1(k).
\end{equation}
The function $\zeta(|k|)$ describes the {\it ultraviolet cutoff} on the wavenumbers $k$. We assume $\zeta$ to be of
class $C^1$ and to have compact support.

The constant $c_{\rm n.o.}$ is
 $$
   c_{\rm n.o.} = [D,D^*] = \frac{2}{\pi}\int_0^\infty
   r|\zeta(r)|^2 {\rm d}r,
 $$
and subtraction of the constant $c_{\rm n.o.}\alpha$ amounts to
normal ordering of the operator $A^2$.

The operator that couples a particle to the magnetic field $B =
{\rm curl}A$ is given by
\begin{equation}\nonumber
\begin{split}
B (x)  = &\displaystyle\sum_{\lambda=1,2}\! \int_{\R^3}\!
\frac{\zeta(|k|)}{2\pi|k|^{1/2}} k\times i\varepsilon_\lambda(k)
\Big[ e^{ikx}\otimes a_\lambda(k)  - e^{-ikx}\otimes
a_\lambda^\ast(k)\Big] \d k
\\
 =: &
K(x) + K^\ast (x).
\end{split}
\end{equation}
In Equation~\eqref{rpf}, $\sigma = (\sigma_1, \sigma_2, \sigma_3)$
is the 3-component vector of Pauli matrices
\begin{eqnarray*}
    \sigma_1 =
    \begin{pmatrix}
      0 & 1 \\
      1 & 0
    \end{pmatrix}\, , \ \
    \sigma_2 =
    \begin{pmatrix}
      0 & -i \\
      i & 0
    \end{pmatrix}\, , \ \
    \sigma_3 =
    \begin{pmatrix}
      1 & 0 \\
      0 & -1
    \end{pmatrix}\, .
\end{eqnarray*}
The photon field energy operator $H_f$ is given by
\begin{equation}\nonumber
H_f = \sum_{\lambda= 1,2} \int_{\R^3} |k| a_\lambda^\ast (k)
a_\lambda (k) \d k.
\end{equation}

The multiplicative potential $W$ is assumed to be short range and in $L^4_{\rm loc}(\R^3)$, and $\lambda$ is a positive
coupling constant. If the negative part of $W$ is nontrivial, then there exists a critical value $\lambda_0$ such that
the Schr\"odinger operator $-\Delta + \lambda W$ has discrete spectrum for all $\lambda>\lambda_0$, but does not have
any discrete spectrum for $0\leq \lambda<\lambda_0$. Analogously, the Pauli-Fierz operator also has a critical coupling
constant $\lambda_1$, which depends on the fine structure constant $\alpha$. It is known \cite{BenguriaVugalter2004}
that $\lambda_1$ converges to $\lambda_0$ from below as $\alpha$ goes to zero.

Before stating our main result, let us introduce some notations. For $v$ a measurable function in $\R^3$, we define
 \begin{equation}\label{def-d1}
  d_v = \frac{1}{2\pi} \left(\int\frac{|v(x)||v(y|}{|x-y|^2}\mathrm{d}x
  \mathrm{d}y\right)^\frac12\, ,
 \end{equation}
if $v$ is not spherically symmetric and
\begin{equation}\label{def-d2}
  d_v=\min\{\frac{1}{2\pi}\left(\int\frac{|v(x)||v(y|}
  {|x-y|^2}\mathrm{d}x \mathrm{d}y\right)^\frac12, \int_0^\infty t |v(t)|
  \mathrm{d}
  t\}
\end{equation}
if $v$ is spherically symmetric.

Our main result is thus
\begin{theorem}\label{mainthm}
Assume that $W(x)$ satisfies the following conditions: $W\in
L^4_{\rm loc}(\R^3)$ and there exists $a>0$, $c>0$  and $\delta>0$
such that for all $|x|>a$, $|W(x)|\leq c(1+|x|)^{-2-\delta}$. Then
 $$
 \lambda_1 \leq \lambda_0 (1 - \alpha \eta^2 +
 \mathcal{O}(\alpha^\frac54))
 $$
with
 $$
  \eta^2 = \frac{1}{6\pi^2} \int_{\R^3}
  \frac{\zeta(|k|)}{|k|(k^2 + |k|+\cw)}
  \mathrm{d} k\, ,
 $$
and
 $$
 C_W = \lambda_0^2 (1+ \lambda_0 d_{W_+}) d_{W^2}\quad\mbox{and}\quad
 W_+ = (|W|+W)/2.
 $$
\end{theorem}


\section{Proof of the main Theorem}

In this section, we will prove the main theorem in the case of
particle with spin $g=1$. The proof for $g=0$ can easily be
deduced with several simplifications.

We start with establishing some useful preliminary estimates.

\subsection{Properties of the self-energy operator T(0) with zero
total momentum}

This section addresses the main properties of the self-energy
operator $T(0)$.
Let us consider the case of a free particle coupled to the
quantized electromagnetic field. The self-energy operator $T$ is
given by
 $$
  T = \left(- i\nabla_{x}\otimes I_f +
  \sqrt{\alpha} A(x)\right)^2 +
  g\sqrt\alpha\sigma \cdot B(x) + I_{el}\otimes H_f
  - c_{\rm n.o.}\alpha.
 $$
We note that this system is translationally invariant, that is,
$T$ commutes with the operator of total momentum
 $$
 P_{tot} = p_{el}\otimes I_f + I_{el}\otimes P_f ,
 $$
where $p_{el}$ and $P_f = \sum_{\lambda =1,2} \int  k
a^\ast_\lambda(k) a_\lambda(k) \d k$ denote the particle and the
photon momentum operators.

Let $\gH_P\cong \C^2\otimes\gF$ denotes the fibre Hilbert space
corresponding to conserved total momentum $P$. For any fixed value
$P$ of the total momentum, the restriction of $T$ to the fibre
space $\gH_P$ is given by (see e.g. \cite{Chen2001})
\begin{equation}T(P) = (P - P_f +
\sqrt{\alpha} A(0))^2 + g \sqrt{\alpha}\sigma\cdot B(0) + H_f -
c_{\rm n.o.}\alpha.
\end{equation}
We denote $\Sigma_0 := \inf\sigma( T(0))$.

For the reader convenience, we first collect in the following
theorem different known facts regarding the ground state of the
operator $T(0)$, which will be used in the proof of the main
theorem.

From now on, we will denote by $\Pi_n$ the projection
 onto the subspace of $\C^2\otimes\gF$ corresponding to vectors which
have all components zero except the $n$-photon components. We also
define $\Pi_n^\geq = 1-\sum_{i=1}^{n-1}\Pi_n$.

For vectors in $\C^2\otimes\gF$, the norm $\|.\|$ will refer to
the standard norm in $\C^2\otimes\gF$.

\begin{theorem}\label{Ogthm}\cite{Frohlich1974, Chen2001, Chenetal2003, Barbarouxetal2003}
For $\a$ sufficiently small we have:
\begin{itemize}
\item $\Sigma_0$ is an eigenvalue bordering to continuous spectrum
of $T(0)$ and $\Sigma_0=\inf\sigma(T)$. \item For any $\Omega_0\in
{\rm Ker}(T(0)-\Sigma_0)$, its projection $\Pi_0\Omega_0$ onto the
zero-photon sector of $\C^2\otimes\gF$ fulfils
$\|\Pi_0\Omega_0\|\neq 0$. If $\Omega_0$ is normalized by
$\|\Pi_0\Omega_0\|=1$, then the following inequalities are
satisfied: $\|\Omega_0\| = 1+ \mathcal{O}(\alpha^{1/2})$, $\|D(0)
\Omega_0 \| = \mathcal{O}(\alpha^{1/2})$, and $\|H_f^{1/2}
\Omega_0 \| = \mathcal{O}(\alpha^{1/2})$. \item For the photon
number operator $N_f:=\sum_{\lambda=1,2}\int \al^*(k)\al(k)\d k$,
we have $\| N_f^{1/2} \Omega_0 \| = \mathcal{O}(\alpha^{1/2})$.
\end{itemize}
\end{theorem}
\begin{corollary}\label{Ogcor} For any vector $\Omega_0\in {\rm Ker}(T(0)-\Sigma_0)$
normalized by $\|\Pi_0\Omega_0\|=1$, we have $\|\Omega_0\| = 1+
\mathcal{O}(\alpha)$, $\|D^*(0)\Pi_1^\geq\Omega_0\| =
\mathcal{O}(\alpha^{1/2})$ and $\|\sigma\cdot
K^*(0)\Pi_1^\geq\Omega_0\| = \mathcal{O}(\alpha^{1/2})$.
\end{corollary}
In the following, we consider two 4-vectors in
$\C^2\otimes\left(L^2(\R^3)\otimes\C^2\right)$, of the form
$\left(\xi(\ua,k,\lambda_1), \xi(\ua,k,\lambda_2),
\xi(\da,k,\lambda_1), \xi(\da,k,\lambda_2)\right)$,  where $\ua$
and $\da$ refer to the spin up and spin down of the particle, and
$\lambda_1$, $\lambda_2$ refer to the two polarizations of the
transverse photons.

$$
   \Gamma_{a,b}:=
   \left(
   \begin{array}{l}
        \Gamma(\ua,k,\lambda_1) \\
        \Gamma(\ua,k,\lambda_2) \\
        \Gamma(\da,k,\lambda_1) \\
        \Gamma(\da,k,\lambda_2)
   \end{array}
   \right)
   :=
   \left(
   \begin{array}{l}
        \frac{\zeta_\Lambda(k)}{|k|^\frac12} (- a \sqrt{k_1^2 + k_2^2} +
        b\, \frac{(k_1 - ik_2)k_3}{\sqrt{k_1^2 + k_2^2}}) \\
        b \zeta_\Lambda(k) \frac{-k_2 - ik_1}
        {\sqrt{k_1^2 + k_2^2}} |k|^\frac12 \\
        \frac{\zeta_\Lambda(k)}{|k|^\frac12} ( b \sqrt{k_1^2 + k_2^2} +
        a \frac{(k_1 + ik_2)k_3}{\sqrt{k_1^2 + k_2^2}}) \\
        a\, \zeta_\Lambda(k)\frac{-k_2 + ik_1}
        {\sqrt{k_1^2 + k_2^2}} |k|^\frac12
   \end{array}
   \right)\, .
 $$
Let
\begin{equation}\label{eq:approx-min}
   \varphi_{a,b} = \sqrt{\alpha}
   \frac{i}
   {2\pi |k|(1 + |k|)}\Gamma_{a,b} .
\end{equation}

\begin{proposition}[Approximate ground state of
the Pauli-Fierz operator]\label{main-prop} For $a$ and $b$ in $\C$
such that $|a|^2+|b|^2=1$, we consider the family of real-valued
functionals $L_{a,b}$ defined on $\C^2\otimes
L^2(\R^3)\otimes\C^2$ by
 $$
  L_{a,b}(\xi) = \bra (k^2 + |k|)\xi,\xi\ket  +
  2\sqrt{\alpha}\Re
    \bra \xi, \Pi_1\sigma\cdot K^*(0)
    (\left(\begin{array}{l}a\\b\end{array}\right)0,0,\cdots)\ket,
 $$
where as before $B(0)=K(0) + K^*(0)$. Then we have

$i)$ The vector $\varphi_{a,b}$ defined by \eqref{eq:approx-min}
is the unique minimizer of $L_{a,b}$.

$ii)$ $|\Sigma_0 - \inf L_{a,b}(\xi)| =\mathcal{O}(\alpha^{3/2})$.

$iii)$ Let $\Omega_0\in {\rm Ker}(T(0)-\Sigma_0)$ be normalized by
$\|\Pi_0\Omega_0\|=1$. Let us denote by $(a,b):= \Pi_0 \Omega_0$.
We define the scalar product $\bra.,.\ket_1$ onto the one-photon
sector $\Pi_1(\C^2\otimes\gF) = \C^2\otimes L^2(\R^3)\otimes\C^2$
by $\bra f,g\ket_1 = \bra (k^2
+|k|)f,g\ket_{\Pi_1(\C^2\otimes\gF)}$. Then for $\gamma\in\R$ and
$R\in\C^2\otimes L^2(\R^3)\otimes\C^2$ such that
 $$
  \Pi_1\Omega_0 = \gamma\varphi_{a,b} + R\
 $$
and $\bra\varphi_{a,b}, R\ket_1 = 0$, we have
\begin{equation}\label{eq:added2}
   \ \bra R,\, R\ket_1 = \mathcal{O}(\alpha^{3/2})\mbox{ and }
   \ |\gamma -1| = \mathcal{O}(\alpha^{3/4})
\end{equation}
\end{proposition}
\begin{remark}
In the above proposition, and in the sequel, we use the same
notation for $\Pi_1\Omega_0$ as a vector in $\C^2\otimes\gF$ which
has all components zero except its one-photon component
$(\Pi_1\Omega_0)^{(1)}$, as well as for the vector
$(\Pi_1\Omega_0)^{(1)}$ in $\C^2\otimes L^2(\R^3)\otimes\C^2$.
\end{remark}

\begin{proof} In this proof, for the sake of simplicity of notations,
we will drop the argument $0$ in the operators $A(0)$, $B(0)$,
$D(0)$, $K(0)$ and their adjoint. We first prove i). Denoting
 $$
  g_{a,b}:=\frac{1}{(k^2+|k|)}\Pi_1\sigma\cdot
  K^*((\begin{array}{l}a
  \\b\end{array}),0,0,\cdots)\, ,
 $$
we have
\begin{equation}\label{eq:lab-other}
 L_{a,b}(\xi) = \bra\xi,\xi\ket_1 + 2\sqrt{\alpha}\Re
 \bra \xi, g_{a,b}\ket_1 = \|\xi + \sqrt{\alpha} g_{a,b}\|_1^2
 - \| \sqrt{\alpha} g_{a,b} \|_1^2 ,
\end{equation}
where $\|.\|_1$ is the norm associated to the scalar product $\bra.,.\ket_1$. Therefore, the minimizer of $L_{a,b}$ is
$-\sqrt{\alpha}g_{a,b}$. A straightforward computation shows that $-\sqrt{\alpha}g_{a,b}= \varphi_{a,b}$. This implies
that
\begin{equation}\label{eq:added1}
  \inf L_{a,b} = L_{a,b}(\varphi_{a,b})
  = - \|\varphi_{a,b}\|_1^2 .
\end{equation}

We now prove ii). We have
 \begin{equation}\label{eq:prop0}
   \begin{split}
      \bra T(0)\Omega_0,\Omega_0\ket = & \bra P_f^2 \Omega_0,
      \Omega_0\ket - \sqrt{\alpha}2\Re \bra P_f\Omega_0, A
      \Omega_0\ket
      + \alpha\bra A^2\Omega_0, \Omega_0\ket\\
      & +\sqrt{\alpha} \bra\sigma\cdot B\Omega_0, \Omega_0\ket
      + \bra H_f\Omega_0, \Omega_0\ket -c_{\rm n.o.}\alpha
   \end{split}
 \end{equation}
Let us estimate the terms in the above equality in order to
identify those who are of order $\alpha^{3/2}$ and higher.
 \begin{equation}\label{eq:prop1}
 \begin{split}
  \bra P_f^2 \Omega_0, \Omega_0\ket  = &
  \bra P_f^2 \Pi_0\Omega_0, \Pi_0\Omega_0\ket
  \!+ \! \bra P_f^2 \Pi_{1}\Omega_0, \Pi_{1}\Omega_0\ket
  \!+ \!\bra P_f^2 \Pi_2^\geq \Omega_0,
  \Pi_2^\geq\Omega_0\ket \\
  = & \bra P_f^2 \Pi_{1}\Omega_0, \Pi_{1}\Omega_0\ket
  \!+ \!\bra P_f^2 \Pi_2^\geq \Omega_0,
  \Pi_2^\geq\Omega_0\ket .
 \end{split}
 \end{equation}
 \begin{equation}\label{eq:prop2}
 \begin{split}
  \bra H_f \Omega_0, \Omega_0\ket = &
  \bra H_f \Pi_0\Omega_0, \Pi_0\Omega_0\ket
  \!+ \! \bra H_f \Pi_{1}\Omega_0, \Pi_{1}\Omega_0\ket
  \!+ \!\bra H_f \Pi_2^\geq \Omega_0,
  \Pi_2^\geq\Omega_0\ket \\
   = & \bra H_f \Pi_{1}\Omega_0, \Pi_{1}\Omega_0\ket
  \!+ \!\bra H_f \Pi_2^\geq \Omega_0,
  \Pi_2^\geq\Omega_0\ket .
 \end{split}
 \end{equation}
Now, using the fact that $n$-photon sectors are invariant under $P_f$, $P_f\Pi_0\Omega_0 =0$, and $\bra P_f \Omega_0,
A\Omega_0\ket = \bra A P_f \Pi_1^\geq\Omega_0, \Omega_0\ket = \bra A \Pi_1^\geq \Omega_0, P_f \Omega_0\ket = \bra A
\Pi_1^\geq \Omega_0, P_f \Pi_1^\geq\Omega_0\ket$, we get
 \begin{eqnarray*}
  |\bra P_f \Omega_0, A\Omega_0\ket| & = &
  |\bra P_f \Pi_1^\geq\Omega_0, A
  \Pi_1^\geq\Omega_0\ket |\\
  &\leq & |\bra P_f \Pi_1^\geq\Omega_0, D
  \Pi_2^\geq\Omega_0\ket |
  + |\bra P_f \Pi_2^\geq\Omega_0, D^*
  \Pi_1^\geq\Omega_0\ket | \\
  & \leq & |\bra P_f\Pi_1\Omega_0, D\Pi_2\Omega_0\ket|
  + |\bra P_f \Pi_2^\geq \Omega_0, D \Pi_3^\geq \Omega_0\ket|\\
  & & + |\bra P_f \Pi_2^\geq \Omega_0, D^* \Pi_1^\geq\Omega_0\ket| \\
  & \leq & \|P_f\Pi_1\Omega_0\|\,\|D\Pi_2\Omega_0\| + \frac12
  \|P_f\Pi_2^\geq\Omega_0\|^2 + \\
  & & 2 \|D \Pi_3^\geq\Omega_0\|^2
  + 2 \|D^*\Pi_1^\geq\Omega_0\|^2 .
 \end{eqnarray*}
Using Theorem~\ref{Ogthm} and Corollary~\ref{Ogcor} and the fact
that $\|P_f\Pi_1\Omega_0\| \leq c(\Lambda) \|\Pi_1\Omega_0\| =
\mathcal{O}(\alpha^{1/2})$, where $c(\Lambda)$ depends only on the
ultraviolet cutoff,  yields
\begin{equation}\label{eq:prop3}
  |\bra P_f \Omega_0, A\Omega_0\ket|  \leq
  \frac12 \|P_f\Pi_2^\geq\Omega_0\|^2 + \mathcal{O}(\alpha) .
\end{equation}
We also have
 \begin{eqnarray*}
  \lefteqn{\bra A^2\Omega_0, \Omega_0\ket} & & \\
  & = & \bra (D+D^*)^2
  \Omega_0, \Omega_0\ket \\
  & = & 2 \Re \bra DD \Omega_0, \Omega_0\ket + 2 \|D
  \Omega_0\|^2 + \|[D, D^*]\|
  \|\Omega_0\|^2 \\
  & = & \mathcal{O}(\alpha^{1/2})+ \mathcal{O}(\alpha) + \|[D,
  D^*]\|(1+\mathcal{O}(\alpha))\, ,
 \end{eqnarray*}
where we used from Theorem~\ref{Ogthm} that $\|\Omega_0\| =
1+\mathcal{O}(\alpha)$ and $\|D\Omega_0\| =
\mathcal{O}(\alpha^{1/2})$. Since the commutator $[D, D^*]$ equals
$c_{\rm n.o.}$ we arrive at
\begin{equation}\label{eq:prop4}
  \bra A^2\Omega_0, \Omega_0\ket = c_{\rm n.o.}+
  \mathcal{O}(\alpha^{1/2}).
\end{equation}
Finally we have, writing $B = K+K^*$
\begin{equation}\nonumber
\begin{split}
  \bra\sigma\cdot B \Omega_0, \Omega_0\ket =
  \bra\sigma\cdot K \Pi_1\Omega_0, \Pi_0\Omega_0\ket
  + \bra\sigma\cdot K\Pi_2^\geq\Omega_0, \Pi_1^\geq \Omega_0\ket \\
   + \bra\sigma\cdot K^*\Pi_0\Omega_0, \Pi_1\Omega_0\ket
  + \bra\sigma\cdot K^* \Pi_1^\geq \Omega_0,
  \Pi_2^\geq\Omega_0\ket\\
   =  2\Re \bra\sigma\cdot K \Pi_1\Omega_0,
  \Pi_0\Omega_0\ket +
  2 \Re \bra\sigma\cdot K^* \Pi_1^\geq \Omega_0,
  \Pi_2^\geq\Omega_0\ket .
\end{split}
\end{equation}
using Theorem~\ref{Ogthm} and Corollary~\ref{Ogcor} we obtain
\begin{equation}\label{eq:prop5}
 \bra\sigma\cdot B \Omega_0, \Omega_0\ket =
 2\Re \bra\sigma\cdot K \Pi_1\Omega_0,
  \Pi_0\Omega_0\ket + \mathcal{O}(\alpha) .
\end{equation}
Collecting \eqref{eq:prop0}-\eqref{eq:prop5} and using $\bra
H_f\Pi_2^\geq\Omega_0, \Pi_2^\geq\Omega_0\ket\geq 0$ we obtain
\begin{equation}
\begin{split}
\bra T(0)\Omega_0, \Omega_0\ket \geq \bra P_f^2\Pi_1\Omega_0,
\Pi_1\Omega_0\ket
 + \bra H_f\Pi_1\Omega_0, \Pi_1\Omega_0\ket
 \\+ 2\sqrt\alpha\Re \bra\sigma\cdot K
\Pi_1\Omega_0,
  \Pi_0\Omega_0\ket + \mathcal{O}(\alpha^\frac32).
\end{split}
\end{equation}
Since on the one-photon sector the operator $P_f^2$ reduces to multiplication by $k^2$, and the operator $H_f$ reduced
to multiplication by $|k|$, we obtain
\begin{equation}\label{eq:ineq1}
\Sigma_0= \frac{\bra T(0)\Omega_0, \Omega_0\ket}{||\Omega_0||^2} \geq L_{a,b}(\Pi_1\Omega_0) +
\mathcal{O}(\alpha^{3/2}) \geq \inf L_{a,b} + \mathcal{O}(\alpha^{3/2}).
\end{equation}
On the other hand, using i), and for
$\psi_{a,b}=(\left(\begin{array}{l}a\\b\end{array}\right),
\varphi_{a,b},0,0,\cdots)$ we have
\begin{equation}\label{eq:ineq2}
 \begin{split}
   \inf L_{a,b} = & L_{a,b}(\varphi_{a,b})
   = \bra T(0) \psi_{a,b}, \psi_{a,b}\ket \geq \Sigma_0
   \|\psi_{a,b}\|^2 \\
   = & \Sigma_0 (1 + \mathcal{O}(\alpha))
   \geq  \Sigma_0 + \mathcal{O}(\alpha^2)
 \end{split}
\end{equation}
Inequalities \eqref{eq:ineq1} and \eqref{eq:ineq2} conclude the
proof of ii).

Eventually, we prove iii). Due to the Inequalities \eqref{eq:ineq1} and \eqref{eq:ineq2}, we have $\inf L_{a,b}
+\mathcal{O}(\alpha^{3/2}) = L_{a,b}(\Pi_1\Omega_0)$. Using \eqref{eq:lab-other}, the fact that $-\sqrt\alpha
g_{a,b}=\varphi_{a,b}$, and \eqref{eq:added1}, we thus get
\begin{equation}
\begin{split}
\inf L_{a,b} +\mathcal{O}(\alpha^{3/2}) = & L_{a,b}(\Pi_1\Omega_0)
= \|\gamma\varphi_{a,b} + R -\varphi_{a,b}\|_1^2 -
\|\varphi_{a,b}\|_1^2 \\
= & (\gamma-1)^2\|\varphi_{a,b}\|_1^2 + \|R\|_1^2 + \inf L_{a,b}\,
,
\end{split}
\end{equation}
which proves \eqref{eq:added2}.
\end{proof}

\subsection{Proof of Theorem~\ref{mainthm}}

As it was mentioned in the introduction, we prove the theorem by constructing a trial function $\Psi$ for which the
quadratic form of $H$ takes a value strictly smaller than $\Sigma_0 \|\Psi\|^2$.

Let us start by proving an auxiliary result. For $\gm\in (0,1)$,
we define $f_\gm\in L^2(\R^3)$ to be a normalized real valued
eigenfunction, with associated eigenvalue $e_\gm$, of the
Schr\"odinger operator
 $$
  h_\gm:= -(1-\gm)\Delta + \lambda_0 W(x).
 $$
Here $\lambda_0$ is the critical coupling constant defined in
Section~\ref{S1}.

\begin{lemma}\label{lem:cw}
Then for $\lambda\leq\lambda_0$, we have
 $$
  \sum_i \!\bra (-\Delta + \lambda W) \frac{\partial f_\gm}{\partial
  x_i}, \frac{\partial f_\gm}{\partial x_i} \ket \!\leq\!
  C_W \| \nabla  f_\gm\|^2 + o_\gamma(1) \|\nabla f_\gm\|^2 ,
 $$
with $C_W := \lambda_0^2 (1+ \lambda_0 d_{W_+}) d_{W^2}$, where $W_+ = (W + |W|)/2$ and $d_{W^2}$ and $d_{W_+}$ are
defined by \eqref{def-d1}-\eqref{def-d2}.
\end{lemma}
\begin{proof}
For a potential $V$ such that $V\in L^2_{\rm loc}$ and short range
we have
 $$
  |\bra V\psi,\psi\ket| \leq d_V \| \nabla \psi\|^2\, .
 $$
Moreover, we know that  $f_\gm$ is an eigenfunction of
$-(1-\gm)\Delta + \lambda_0 W$ and that the associated eigenvalue
$e_\gm$ tends to zero as $\gamma$ tends to zero, since $\lambda_0$
is the critical coupling constant. Therefore we obtain the
following sequence of inequalities
\begin{equation}\label{eq:19bis}
\begin{split}
\sum_{i}\bra (-\Delta + \lambda W) \frac{\partial f_\gm}{\partial
x_i}, \frac{\partial f_\gm}{\partial x_i}\ket \leq \sum_{i}\bra
(-\Delta + \lambda W_+) \frac{\partial
f_\gm}{\partial x_i}, \frac{\partial f_\gm}{\partial x_i}\ket\\
\leq \sum_{i}\bra (-\Delta + \lambda_0 W_+) \frac{\partial
f_\gm}{\partial x_i}, \frac{\partial f_\gm}{\partial x_i}\ket \leq
 \sum_i \bra -\Delta  \frac{\partial f_\gm}{\partial x_i},
\frac{\partial f_\gm}{\partial x_i}\ket (1+d_{\lambda_0 W_+}) \\
=  (1+d_{\lambda_0 W_+})  \bra -\Delta f_\gm, -\Delta f_\gamma\ket
= (1+d_{\lambda_0 W_+}) \bra \frac{e_\gm  - \lambda_0 W }
{1-\gamma}f_\gm, -\Delta f_\gm\ket \\
= \frac{e_\gm(1+d_{\lambda_0 W_+})}{1-\gamma} \bra f_\gm, -\Delta
f_\gm\ket
 - \frac{\lambda_0(1+d_{\lambda_0
W_+})}{1-\gamma} \bra W f_\gm, -\Delta f_\gm\ket
\end{split}
\end{equation}
We estimate the last term in the right hand side by
\begin{equation}\label{eq:20bis}
\begin{split}
 - \bra W f_\gm, -\Delta f_\gm\ket =  \bra W f_\gm,
 \frac{\lambda_0 W - e_\gm}
{1-\gamma} f_\gm\ket \\
\leq -\frac{e_\gm d_{W_+}}{1-\gamma} \|\nabla f_\gm\|^2 +
\frac{\lambda_0}{1-\gamma} d_{W^2} \|\nabla f_\gm\|^2
\end{split}
\end{equation}
The Inequalities \eqref{eq:19bis} and \eqref{eq:20bis} imply for
$\lambda\leq\lambda_0$
\begin{equation}
 \sum_{i}\bra (-\Delta + \lambda W) \frac{\partial
f_\gm}{\partial x_i}, \frac{\partial f_\gm}{\partial x_i}\ket \leq
\left(\lambda_0^2 (1+\lambda_0 d_{W_+}) d_{W^2} +
o_{\gm}(1)\right) \|\nabla f_\gm\|^2
\end{equation}
\end{proof}
In the rest of this section, we will mainly work in the space representation for both particle and photons. Following
\cite{Chenetal2003}, let us introduce, for given $x\in\R^3$, the shift operator on the photon space variables
$\tau_x:\C^2\otimes\gF\rightarrow\C^2\otimes\gF$. For $\phi=(\phi_0, \phi_1, \ldots, \phi_n, \ldots )\in
\C^2\otimes\mathcal{F}$, we have, writing by abuse of notation $\tau_x\phi = (\tau_x\phi_0 ,\tau_x\phi_1 ,\ldots)$,
 $$
   \tau_x\phi_n (s; y_1, \ldots, y_n; \lambda_1, \ldots,
   \lambda_n) = \phi_n (s; y_1 - x, \ldots, y_n - x; \lambda_1, \ldots,
   \lambda_n) ,
 $$
where $s$ is the spin of the particle and takes value in $\{
\uparrow,\, \downarrow\}$.

We denote by $\Omega^x_0$ the ground state $\Omega_0$ written in
space representation and shifted by $x$, i.e.
 $$
  \Omega_0^x := \tau_x \mathcal{F}^{-1}\Omega_0,
 $$
where $\mathcal{F}$ stands for the Fourier transform.

Recall that $D^*(0)$ is an operator valued vector with $3$
components which we denote by $D^*(0)_i$ ($i=1,2,3$). Then we
consider the functions
\begin{equation}
 \theta_{i} = (0,\theta_{i}^{(1)},0,\ldots)
 \in\C^2\otimes\gF
\end{equation}
with
\begin{equation}\label{eq:def2-theta}
\theta_{i}^{(1)} = (k^2+|k|+\cw)^{-1} \Pi_1 D^*(0)_i \left(
   (\begin{array}{l}
   a \\ b
   \end{array}),0,\ldots
   \right)\,
\end{equation}
and
\begin{equation}
 \theta_{i}^x = \tau_x
 \mathcal{F}^{-1}\theta_{i}.
\end{equation}
We first state some properties of $\theta_{i}$.
\begin{lemma}\label{lem-theta}$\ $

\noindent i) For $i\neq j$ we have
 $$
 \bra \theta_{i}, \theta_{j} \ket = 0
 \quad \mbox{and} \quad \bra \theta_{i},
 \theta_{j} \ket_1 = 0 .
 $$
ii) For $i=1,2,3$ holds
 $$
  \|\theta_{i}\sqrt{k^2 + |k|+\cw}\|^2 = \frac{1}{6\pi^2} \int_{\R^3}
  \frac{\zeta(|k|)}{|k|(k^2 + |k|+\cw)}
  \mathrm{d} k,
 $$
iii) For $i=1,2,3$, $\bra k_i \varphi_{a,b}, \Pi_1\theta_{i}\ket =
0$.
\end{lemma}
\begin{proof}
To prove this Lemma we remind that $\theta_{i}$ has only a non
zero component $\Pi_1 \theta_{i}$ in the one photon sector, and
 $$
  \Pi_1 \theta_{i}
  = \left(
  \begin{array}{l}
    a \frac{\varepsilon_{1,i}(k) \zeta(|k|)}
    {|k|^\frac12 (k^2 + |k|+\cw)} \vspace{0.1cm}\\
    a \frac{\varepsilon_{2,i}(k) \zeta(|k|)}
    {|k|^\frac12 (k^2 + |k|+\cw)} \vspace{0.1cm} \\
    b \frac{\varepsilon_{1,i}(k) \zeta(|k|)}
    {|k|^\frac12 (k^2 + |k|+\cw)} \vspace{0.1cm}\\
    b \frac{\varepsilon_{2,i}(k) \zeta(|k|)}
    {|k|^\frac12 (k^2 + |k|+\cw)}\
 \end{array}
 \right),
 $$
where the two polarization vectors $\varepsilon_1(k)$ and
$\varepsilon_2(k)$ are defined in \eqref{def-eps}. The properties
stated in the Lemma follow straightforwardly from computations of
the corresponding integrals.
\end{proof}

We consider the trial function $\Psi\in
L^2(\R^3)\otimes\C^2\otimes\gF$:
\begin{equation}
 \Psi := \Psi_1 + \Psi_2:= f_\gm(x) \Omega_0^x
  + i\sqrt{\alpha} \sum_{i=1}^3
 \theta_{i}^x \frac{\partial f_\gm(x)}{\partial x_i}\, .
\end{equation}

Now we compute the expectation value of $H$ in the state $\Psi$.
We have
 $$
  \bra H\Psi,\Psi\ket = \bra H\Psi_1,\Psi_1\ket +
  \bra H\Psi_2,\Psi_2\ket + 2\Re \bra H\Psi_1,\Psi_2\ket.
 $$
As usual \cite{Chenetal2003}, due to the orthogonality $\bra
f,\partial f/\partial x_i\ket =0$, we have
\begin{equation}\label{eq:22}
\bra H\Psi_1, \Psi_1\ket = \Sigma_0\|\Psi_1\|^2 +  \bra (-\Delta +
\lambda W(x))f_\gm,f_\gm\ket
 \|\Omega_0\|^2
\end{equation}

Since $\Psi_2$ has only a non zero component in the one photon
sector, in the quadratic form $\bra H \Psi_2,\Psi_2\ket$, all the
terms involving $A(0)$ or $B(0)$ vanish. Moreover, using
Lemma~\ref{lem-theta} and the orthogonalities $\bra \frac{\partial
f_\gm}{\partial x_i}, \frac{\partial f_\gm}{\partial x_j}\ket =0$
and $\bra \frac{\partial f_\gm}{\partial x_i}, \frac{\partial^2
f_\gm}{\partial x_i\partial x_j}\ket =0$, for $i\neq j$, we arrive
at
\begin{equation}\label{eq:23}
\begin{split}
 \bra H \Psi_2, \Psi_2\ket = & \alpha  \sum_{l}
 \|\theta_{l}^x\|^2 \bra (-\Delta+\lambda
 W)\frac{\partial f_\gm}{\partial x_l},
 \frac{\partial f_\gm}{\partial x_l}\ket \! + \!
 \mathcal{O}(\alpha^2) \|\nabla f_\gm\|^2 \\ &
 + \alpha
 \sum_{l}\|\frac{\partial f_\gm}{\partial x_l}\|^2
 \bra (|k|+k^2) \Pi_1\theta_{l}^{x},
 \Pi_1\theta_{l}^{x}\ket
 .
\end{split}
\end{equation}
To compute the last term $\bra H\Psi_1,\Psi_2\ket$, we first note
that
\begin{equation}\nonumber\label{eq:cross-zero}
\begin{split}
\bra (-\Delta + \lambda W) f_\gm, \frac{\partial f_\gm}{\partial
x_i}\ket =&
 \bra (-(1-\gm)\Delta + \lambda W) f_\gm, \frac{\partial f_\gm}{\partial
x_i}\ket -\gm \bra \Delta  f_\gm, \frac{\partial f_\gm}{\partial
x_i}\ket \\
=& e_\gm \bra f_\gm, \frac{\partial f_\gm}{\partial x_i}\ket +
\gamma \sum_j \bra \frac{\partial^2 f_\gm}{\partial x_j^2},
\frac{\partial f_\gm}{\partial x_i}\ket =0 .
\end{split}
 \end{equation}
The last equality holds since $f_\gm$ is a real function vanishing
at infinity. Moreover, all other terms in the quadratic form $\bra
H\Psi_1,\Psi_2\ket$ which contain $\bra f_\gm, \frac{\partial
f_\gm}{\partial x_i}\ket$ vanish also. So we arrive at
\begin{equation}\label{eq:24}
 \begin{split}
   2\Re \bra H\Psi_1,\Psi_2\ket = & - 2\Re \bra
   P\cdot
   (P_f-\sqrt{\alpha}A(0))\Psi_1, \Psi_2\ket \\
   = & 2 \sqrt{\alpha} \sum_i \|\frac{\partial f_\gm}
   {\partial x_i}\|^2
   \Re \bra (P_f- \sqrt{\alpha} A(0))_i\Omega_0,
   \theta_{i}\ket
 \end{split}
\end{equation}
The term with $P_f$ on the right hand side  is estimated as
follows
\begin{equation}\label{eq:above}
 \begin{split}
    \Re \bra (P_f)_i \Omega_0,
    \theta_{i}\ket_{\C^2\otimes\gF}
    = & \Re \bra
    \Pi_1(P_f)_i(\gamma\varphi_{a,b} + R),
    \Pi_1\theta_{i}\ket_{\C^2\otimes L^2(\R^3)\otimes\C^2}\\
    = & \Re \bra
    k_i(\gamma\varphi_{a,b} + R), \Pi_1\theta_{i}
    \ket_{\C^2\otimes L^2(\R^3)\otimes\C^2}\\
    \leq & \|R\|_1\||k|^\frac12 \Pi_1\theta_{i}\|
    + \gamma \Re \bra
    k_i\varphi_{a,b}, \Pi_1\theta_{i}
    \ket_{\C^2\otimes L^2(\R^3)\otimes\C^2}
 \end{split}
\end{equation}
Using Proposition~\ref{main-prop} yields the following bound for
the first term in the right hand side of \eqref{eq:above}
\begin{equation}\label{eq:26}
  \|R\|_1\||k|^\frac12 \Pi_1\theta_{i}\| =
  \mathcal{O}(\alpha^{\frac34})
\end{equation}
According to Lemma~\ref{lem-theta}~iii), the second term in the
right hand side of \eqref{eq:above} equals zero. Therefore,
collecting \eqref{eq:24}-\eqref{eq:26}, we arrive at
\begin{equation}\label{eq:27}
 2\Re\bra H \Psi_1, \Psi_2\ket =
  - 2 \alpha \sum_i \|\frac{\partial f_\gm}
   {\partial x_i}\|^2
   \Re \bra A(0)_i\Omega_0,
   \theta_{i}\ket +
   \mathcal{O}(\alpha^{\frac54}) \|\nabla f_\gm\|^2
\end{equation}
Now we have, using Theorem~\ref{Ogthm} and the fact that $D(0)$
restricted to the 2-photon sector is a bounded operator
\begin{equation}\label{eq:28}
 \begin{split}
   \Re \bra A(0)_i\Omega_0,
   \theta_{i}\ket = &\Re \bra D(0)_i\Pi_2\Omega_0,
   \theta_{i}\ket
   + \Re \bra D^*(0)_i\Pi_0\Omega_0, \theta_{i}\ket\\
   & = \mathcal{O}(\alpha^{\frac12}) +
   \Re \bra D^*(0)_i\Pi_0\Omega_0, \theta_{i}\ket .
 \end{split}
\end{equation}
Due to the definition~\eqref{eq:def2-theta} of $\theta_{i}$, the second term on the right hand side of \eqref{eq:28} is
$\| \theta_{i} \sqrt{k^2 +|k| + C_W}\|^2$. Therefore, collecting the Equalities~\eqref{eq:22}, \eqref{eq:23},
\eqref{eq:27} and \eqref{eq:28} we obtain
\begin{equation}
 \begin{split}
   \bra H\Psi, \Psi\ket = & \Sigma_0\|\Psi_1\|^2 +
 \|\Omega_0\|^2
 \bra (-\Delta + \lambda W)f_\gm,f_\gm\ket \\
 & + \alpha \sum_{l}
 \|\theta_{l}\|^2 \bra (-\Delta+\lambda
 W)\frac{\partial f_\gm}{\partial x_l},
 \frac{\partial f_\gm}{\partial x_l}\ket
 \\ & - 2\alpha
 \sum_{l}\|\frac{\partial f_\gm}{\partial x_l}\|^2
 \| \theta_{i} \sqrt{k^2 +|k| + C_W}\|^2
 + \mathcal{O}(\alpha^{\frac54}) \|\nabla f_\gm\|^2
 \\ & + \alpha  \sum_{l}\|\frac{\partial f_\gm}{\partial x_l}\|^2
 \|\theta_{l}\|_1^2 .
 \end{split}
\end{equation}
From Lemma~\ref{lem-theta}, we know that $\|\theta_l \sqrt{k^2
+|k| + C_W}\|^2$ is independent of $l$. We denote this constant by
$\eta^2$. With Lemma~\ref{lem:cw} we thus arrive at
\begin{equation}\label{eq:last1}
\begin{split}
 \bra H\Psi, \Psi\ket \leq \Sigma_0\|\Psi\|^2 -\Sigma_0\|\Psi_2\|^2 +
 \|\Omega_0\|^2\left( \|\nabla f_\gm\|^2 +
 \bra \lambda W f_\gm,f_\gm\ket\right)\\
 - \alpha \sum_l \|\frac{\partial f_\gm}{\partial x_l}\|^2
  \eta^2
 + \alpha o_\gamma(1)\|\nabla f_\gm\|^2 + \mathcal{O}(\alpha^{\frac54})
 \|\nabla f_\gm\|^2\, .
\end{split}
\end{equation}
Note that $\Sigma_0 \|\psi_2\|^2 = \mathcal{O}(\alpha^2) \|\nabla
f_\gm\|^2$. We thus obtain
\begin{equation}\label{eq:last2}
\begin{split}
 \bra H\Psi, \Psi\ket - \Sigma_0\|\Psi\|^2 \leq   \\
 \|\Omega_0\|^2 \Big((1-\frac{\alpha}{\|\Omega_0\|^2}
 \eta^2 +
 \frac{\alpha}{\|\Omega_0\|^2} (o_\gm(1)+ \mathcal{O}(\alpha^\frac14)))\|\nabla
 f_\gm\|^2 +
 \bra \lambda W f_\gm,f_\gm\ket \Big)
\end{split}
\end{equation}
Using from Corollary~\ref{Ogcor} that $\|\Omega_0\|^2 =
1+\mathcal{O}(\alpha)$, we obtain
\begin{equation}\label{eq:last3}
 \begin{split}
  \bra H\Psi, \Psi\ket - \Sigma_0\|\Psi\|^2 \leq   \\
  \|\Omega_0\|^2 \Big( (1-\alpha \eta^2 +
  \alpha o_\gm(1)+ \mathcal{O}(\alpha^\frac54))\|\nabla
 f_\gm\|^2 +
  \bra \lambda W f_\gm,f_\gm\ket \Big)\, .
 \end{split}
\end{equation}
Therefore
\begin{equation}\label{eq:last4}
 \begin{split}
   \bra H\Psi, \Psi\ket - \Sigma_0\|\Psi\|^2 \leq
   \|\Omega_0\|^2 (1-\alpha \eta^2 +
   \alpha o_\gm(1)+ \mathcal{O}(\alpha^\frac54))
   \\ \times \Big( \|\nabla
   f_\gm\|^2 + (1+\alpha \eta^2 +
   \alpha o_\gm(1)+ \mathcal{O}(\alpha^\frac54))^{-1}
   \bra \lambda W f_\gm,f_\gm\ket \Big)\, .
 \end{split}
\end{equation}
If $\lambda > \lambda_0 (1 - \alpha \eta^2 +
\mathcal{O}(\alpha^\frac54))$, choosing $\gamma$ (depending on
$\alpha$) small enough, we arrive at $\bra H\Psi,\Psi\ket -
\Sigma_0\|\Psi\|^2 <0$. Due to Griesemer et al.
\cite{Griesemeretal2001}, this implies the existence of a ground
state for $H$.


\bibliographystyle{plain}

\end{document}